\documentclass[twocolumn,showpacs, preprintnumbers,  ]{revtex4}
\usepackage{amssymb}

\usepackage{graphicx}
\usepackage{dcolumn}
\usepackage{bm}

\begin{document}

\preprint{APS/123-QED}
\title{Pseudogap behavior of nuclear spin relaxation in high $T_c$
superconductors in terms of phase separation}
\author{L.\ P.\ Gor'kov}
\address{National High Magnetic Field Laboratory, Florida State University,
Tallahassee, FL 32310, USA\\
and L.D.Landau
Institute for Theoretical Physics, 142432 Chernogolovka, Russia}
\author{G.\ B.\ Teitel'baum,\thanks{
E-mail: grteit@dionis.kfti.knc.ru}}
\address{E.K.Zavoiskii Institute for Technical Physics of the RAS,
420029 Kazan, Russia}

\date{\today }

\begin{abstract}
We analyze anew experiments on the NMR in cuprates and find an
important information on their phase separation and its stripe
character hidden in the dependence of $1/^{63}T_{1}$ on degree of
doping.
 In a broad class of materials $1/^{63}T_{1}$ is the sum of
two terms: the temperature independent one attributed to
``incommensurate'' stripes that occur at external doping, and an
``universal'' temperature dependent term ascribed to moving
metallic and AF sub-phases. We argue that the frustrated first
order phase transition in a broad temperature interval bears a
dynamical character.
\end{abstract}
 \maketitle

\address{E.K.Zavoiskii Institute for Technical Physics of the RAS,\\
Sibirskii Trakt 10/7,\\
Kazan 420029, RUSSIA}

\address{NHMFL, 1800 E P.Dirac Dr., \\
Tallahassee FL 32310, USA}

\address{E.K.Zavoiskii Institute for Technical Physics of the RAS,\\
Sibirskii Trakt 10/7,\\
Kazan 420029, RUSSIA}

\address{NHMFL, 1800 E P.Dirac Dr., \\
Tallahassee FL 32310, USA}


%
Soon after the discovery of high temperature superconductivity (HTSC)
in cuprates by Bednorz and Muller in 1986 it was suggested theoretically %
\cite{1} that the materials should manifest a tendency to inherent phase
segregation involving both lattice and electronic degrees of freedom. Later
other models revealed similar tendency to an inhomogeneous electronic ground
state \cite{2,3,4,5}. (For an experimental summary see e.g., \cite{6,8,
Howald}). It is currently common to discuss microscopic phase separation
(PS) in cuprates as a pattern of the alternating spin-charge stripes \cite%
{Salcola}, dynamical or pinned by static defects seen, e.g., in (LaNd)$%
_{2-x} $Sr$_{x}$CuO$_{4}$ \cite{9} with Nd doping.

The most intriguing phenomenon in the normal state of cuprates is the
pseudogap (PG)  seen  as a new energy scale or a crossover temperature, $%
T^{\ast }$, in NMR, tunneling spectra, resistivity etc. (for review see,
e.g., \cite{14,15}). We address this highly debated issue by considering the
PG regime as a regime of the phase segregation.

Our premise is that $T^{\ast }(x)$ marks a temperature below which
the system enters mesoscopic dynamical regime of a frustrated 1st
order phase
transition with sizes of the sub-phases` determined by electroneutrality %
\cite{1,17}. Unlike \cite{Chakra, Kam}, changes in the normal properties
below such a $T^{\ast }$ come not from a symmetry breaking but due to
interweaving of regions with different holes content.

The NMR fingerprints of a static stripe phase were observed for the low
temperature tetragonal (LTT) compounds (LaEu)$_{2-x}$Sr$_{x}$CuO$_{4}$ \cite%
{18} and (LaNd)$_{2-x}$Sr$_{x}$CuO$_{4}$ \cite{19}. Dynamical features are
seen in inelastic neutron experiments at frequencies 10$^{12}$-10$^{13}$ sec$%
^{-1}$, exceeding the NMR scale.

In the present paper we analyze anew vast \textit{experimental}
data on nuclear spin
relaxation in cuprates to show that for a broad class of materials $%
^{63}T_{1}^{-1}$ in the PG regime is the \textit{sum of the two
terms}: the temperature independent one that we attribute to
stripes caused by the presence of dopants and an ''universal''
temperature dependent term related to the moving metallic and
antiferromagnetic (AF) subphases. We argue that for LSCO $T^{\ast
}$ is above the ''dome'' of superconductivity (SC).

For a guidance we adopt a qualitative two-band view \cite{1,17}. Basic to it
is the assumption that in absence of interactions the energy spectrum of
holes for CuO$_{2}$ plane consists of an itinerant band and the periodic
array of local levels identified with the $d^{9}$ Cu-configuration. Although
hole concentration is not a thermodynamic variable, we suggest that  doping
is accompanied by changes in  the charge transfer gap and implements a 1-st
order Mott transition. (The low temperature Hall effect  \cite{31,33,34}
show that the number of the itinerant holes rapidly increases up to one per
unit cell near $x=0.15$).

New features in \cite{1,17} as compared to other models \cite{2,3,4,5}
concern Cu spins: 1) a hole on a $d^{9}$ Cu site is bound by the lattice
(remnants of the Jahn-Teller effect); 2) a local hole on a $d^{9}$ Cu-site
distorts surrounding lattice so that the distortions on different sites
interact with each other. The elastic energy minimizes itself by bringing
occupied Cu sites together, thus forming nuclei of a dense ``liquid`` phase
of local centers` (in the language of the lattice ``liquid-gas''). It is the
\textit{lattice} that triggers ``liquid-gas'' transition! An exchange
between holes' spins on the neighboring sites tends to organize spins of
``liquid'' AF sub-phase.

For holes localized on Cu`s in the CuO$_{2}$ plane the lattice
``liquid-gas'' transition at some $T^{\ast }$ is well-known from the exact
solution of the familiar 2D Ising-problem. Taking hole-doped \textrm{La}$%
_{1-x}$\textrm{Sr}$_{x}$\textrm{CuO}$_{4}$, for the sake of argument,
missing (with respect to the Sr ionicity) holes` density in the AF area must
be recompensated by local metallic inclusions or droplets with higher hole's
content.

We speculate that tiny structure of sub-phases leads to strong fluctuations
in islands` sizes and positions \cite{1}. Boundaries between them move
rapidly. Moving boundary itself is due to the new notion that a Cu-site can
merely loose spins when a ``metallic'' regions crosses over the position of
that $^{63}$Cu nuclear spin. Experimentally one sees only one nuclear
resonance frequency, this provides a strict evidence in favor of the
dynamical picture.

In what follows we address only $1/^{63}T_{1}$ behaviour because for
cuprates AF fluctuations prevail over the Korringa mechanisms.

We now turn to experimental data on $1/^{63}T_{1}$ for a number of
cuprates. At high enough temperatures itinerant holes and Cu
$d^{9}$ states form a homogeneous phase. In particular we will
also address means to determine $T^{\ast }$, which because of the
Coulomb effects manifests itself only as a start of a new
fluctuation regime.
\begin{figure}[h]
\includegraphics[width= 8 cm]{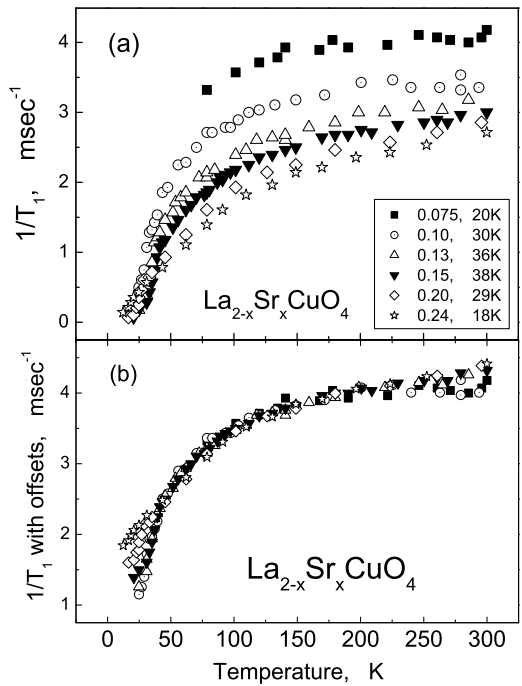}
\caption{The temperature dependence of $1/^{63}T_{1}(x)$ for LSCO: a) the
plots for different $x$ and $T_{c}$ (see inset) are taken from \protect\cite%
{39}, at higher temperatures all of them converge to the same
value of 2.7 msec$^{-1}$ \protect\cite{Imai} ; b) the same
dependencies collapsing to the single curve after the
corresponding vertical offsets.}
\label{Fig1}
\end{figure}

A widespread view is that PG $T^{\ast }(x)$ is a density of states (DOS)
crossover. In the $(T,x)$ plane it crosses  the SC ''dome'', $T_{c}$ (see
e.g. in \cite{Schmallian, 27}). We take a different view. In Fig. 1a we
collected data on $1/^{63}T_{1}$ in LSCO from \cite{39}. The following
comments are of relevance here: 1) according to \cite{Imai} $1/^{63}T_{1}(T)$
at higher temperatures tends to 2.7 msec$^{-1}$ \textit{for all} Sr
concentrations, in spite of considerable spread seen in Fig. 1a. Beginning
of deviation from that value could be  a definition of $T^{\ast }(x)$; 2)
note that $1/^{63}T_{1}$ i.e. dissipation, monotonically decreases with the
decrease of disorder from small $x$ to 0.24; 3) after an appropriate \textit{%
vertical} offset all curves in Fig. 1a collapse onto the $T$ dependence of $%
1/^{63}T_{1}$ for the ``optimal'' $x=0.15$ above 50 K (Fig. 1b). Same
tendency is seen in Fig. 2a,b for YBCO (6.5) doped with Ca: the data in Fig.
2a for different $z$ in Y$_{1-z}$Ca$_{z}$Ba$_{2}$ Cu$_{3}$ O$_{6.5}$ \cite%
{Singer} all fall on the top of each other in Fig. 2b.

\begin{figure}[h]
\includegraphics[width= 8 cm]{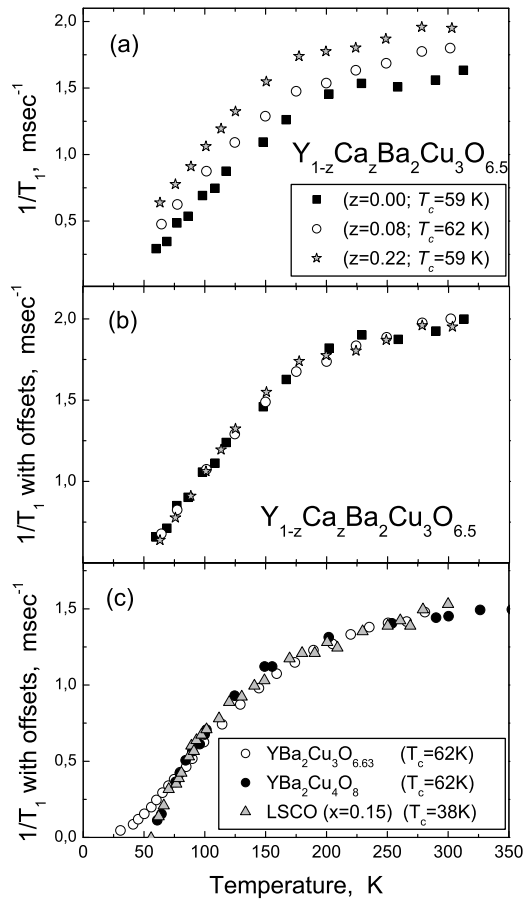}
\caption{The temperature dependence of $1/^{63}T_{1}$ for different
compounds: a) for Y$_{1-z}$Ca$_{z}$Ba$_{2}$Cu$_{3}$O$_{6.5}$ with different
Ca content $z$ \protect\cite{Singer} ; b) the same curves overlayed at each
other by the vertical offsets; c) the $1/^{63}T_{1}$ for YBCO (123) %
\protect\cite{26} overlayed with that for LSCO \protect\cite{39} and YBCO
(124) \protect\cite{ 27}.}
\label{Fig2}
\end{figure}

This prompts us to verify whether same ''off-settings'' of the $1/^{63}T_{1}$
data apply to a broader group of materials. The stoichiometric YBa$_{2}$Cu$%
_{4}$O$_{8}$ possesses no structural or defect disorder and we adjust all
data  to the $1/^{63}T_{1}$ behaviour for this material \cite{27}. Fig. 2c
shows that after a vertical shift in $1/^{63}T_{1}$ all the materials indeed
follow the same ''universal'' temperature dependence above their $T_{c}$ and
below 300 K. In other words, in this temperature range the nuclear spin
relaxation in these cuprates is a sum of contributions from two parallel
processes:
\begin{equation}
1/^{63}T_{1}=1/^{63}\overline{T}_{1}(x)+1/^{63}\widetilde{T}_{1}(T)
\end{equation}%
In eq. (1) $1/^{63}\overline{T}_{1}(x)$ depends on a material and a degree
of disorder $(x)$, but does not depend on temperature, while $1/^{63}%
\widetilde{T}_{1}(T)$, depends only on temperature, is the same for all
these compounds and coincides with the $1/^{63}T_{1}$ for the two chains
YBCO 124 above its $T_{c}$=62 K.

The decomposition (1) into two parallel dissipation processes show that
usual definitions of $T^{\ast }$ \cite{Schmallian} have no grounds. In
Fig.1a the LSCO data with $x<0.15$ are spread even above 250 K. As a rough
estimate for  $T^{\ast }$, it  is much higher than the  SC onset temperature.

Fig. 3 presents the dependence on $x$ for $1/^{63}\overline{T}_{1}$ in La$%
_{2-x}$Sr$_{x}$CuO$_{4}.$ The inset provides the ''offsets'' (i.e. $1/^{63}%
\overline{T}_{1}$- terms) for other materials. We return to
discussion of the two terms of eq. (1) later.

The observation that is central for the following is that in all the
materials with non-zero $1/^{63}\overline{T}_{1}$ incommensurate (IC) peaks
have been observed in  neutron scattering \cite{G}. Peaks are close to the $%
[\pi ,\pi ]$ -- point: at $[\pi (1\pm \delta ),\pi ]$ and $[\pi ,\pi (1\pm
\delta )]$ \cite{9}. We will now look for the connection between these two
phenomena.

Discovery of IC spin fluctuations presented a challenge for explaining the
NMR results for the oxygen spin relaxation times: hyperfine field ''leaks''
originated by the AF incommensurate fluctuations, would considerably
increase the oxygen's relaxation rates, but this was not seen
experimentally. Slichter (see in Ref.\cite{Barzykin}) interpreted these
contradictions in terms of ''discommensurations'': a periodic array of
soliton-like walls separating regions with a short-range AF order. Unlike
neutrons, the NMR as a local probe, does not feel the overall periodicity .

\begin{figure}[h]
\includegraphics[width= 8 cm]{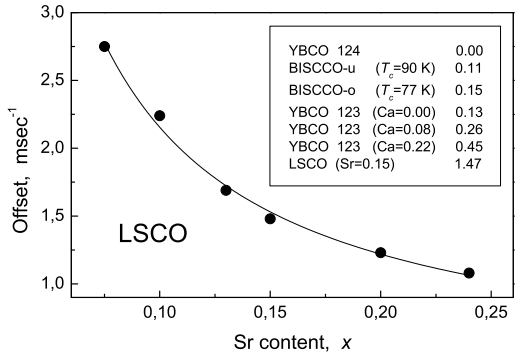}
\caption{The offset $\ 1/^{63}\overline{T}_{1}(x)$ vs Sr content
$x$ for LSCO (relative to that for YBCO 124), line is a guide for
eyes. Inset: the offsets for some other compounds (data for
underdoped (u) and overdoped (o) BISCCO 2212 deduced from
\protect\cite{Walstedt}; to compare BISCCO with LSCO and YBCO
materials the hyperfine constants have to be properly adjusted).}
\label{Fig3}
\end{figure}

As a theory attempt to put the IC peaks into the context of the \textit{%
dynamical} PS separation, consider first a stoichiometric material
like YBCO 124. Below some $T^{\ast }$ we expect the system to
break-up into sub-phases -- the one with a richer content of the
Cu-spins (``AF''-phase), and the second - with an excess of
carriers \cite{1,17}. The system passes $T^{\ast }$ gradually and
remains macroscopically homogeneous. PS would
express itself in strong fluctuations that provide for the $1/^{63}%
\widetilde{T}_{1}(T)$ - term in eq. (1).

At doping the system (LSCO) must screen excess charge (Sr$^{2+}$) in AF
regions. External doping introduces features that may have something to do
with pinning or structural changes. For the Nd doped LSCO, La$_{2-x-y}$Nd$%
_{y}$Sr$_{x}$CuO$_{4}$ (for the summary below see experiments: \cite%
{Crawford, Tranquada1, Tranquada2, Tranquada3}), \ doping of La$_{1.6}$Nd$%
_{0.4}$CuO$_{4}$ with Sr reproduces all features, including SC at lower
temperatures, of La$_{1-x}$Sr$_{x}$CuO$_{4}$ itself with that important
difference that ''stripes'' easily become static. \textit{Elastic
lattice/charge} IC peaks were seen in \cite{Tranquada2} at $\pi \lbrack 2\pm
\delta ,0,0]$ with the lattice distortions along the modulation direction
(IC peaks positions in \cite{Tranquada2} are defined by $\epsilon =\delta /2$%
). Lattice peaks appear first at cooling, followed by the static magnetic
peaks. According to \cite{Tranquada2}, pinning of IC distortions can be
ascribed to the pattern of the octahedra tilts and redistribution of the
Cu-O - bonds in the LTT phase that sets in LaNdSCO at higher temperatures %
\cite{Crawford}. (Unlike the high temperature tetragonal phase (HTT) the LTT
phase contains four formula units per primitive cell). Strips of AF ordered
phase alternate with ''metallic'' domain walls. Such strip arrangement by
itself is nothing but an optimization of the competing Coulomb and lattice
forces \cite{3}.

Note that the stripes at low temperature finally acquire  a long range order
even in LSCO \ (at smaller $x$ \cite{Fujita}), breaking the symmetry of the
ground state. The way of the ''coexistence'' of SC and the stripe order in
the same sample remains unresolved: one view treats the new stripe symmetry
as a superstructure superimposed on the  Fermi surface that  change the
energy spectrum like any SDW/CDW can do it (e.g. \cite{Salcola}). Another
plausible alternative would be a spatially inhomogeneous coexistence of the
nonsuperconducting IC AF phase  and a ''metallic'' phase with strong
fluctuations.

Coexistence of a SC and the IC AF phases at low temperatures was confirmed
recently by the neutron diffraction experiments \cite{Lake} for La$_{2-x}$Sr$%
_{x}$CuO$_{4}$ $(x=0.10)$ in the vortex state. (The coexistence of SC and AF
formations was found also from the $\mu $SR spectra \cite{Niedermeier}).

One can now say that with the temperature increase the static phase
coexistence melts into the dynamical regime governed by the Coulomb
interactions. ``Discommensurations'' suggested by Slichter (see in \cite%
{Barzykin}) ought to be corrected: IC excitations should indeed consist of
alternating strips of the short-range AF order, but separated by walls that
store the compensating charge. Direct correspondence between $x$ and $\delta
$: $\delta /2$ $(\equiv \epsilon )\approx x$ found for $0.05$ $\leq x\leq
1/8 $ saturates for $x>1/8$ at $\delta \approx 1/4$ \cite{Yamada}. Fig.3
allows to consider the tendency as an increase in the share of the
''metallic'' fraction.

We shall now make an attempt to agree on a semi-quantitative level the
observed IC magnetic peaks in La$_{2-x}$Sr$_{x}$CuO$_{4}$ with the values of
the first term in eq. (1). We concentrate on La$_{1.86}$Sr$_{0.14}$CuO$_{4}$
for which the most detailed data are available \cite{Aeppli}.

With the notation from \cite{Zha}
\begin{equation}
1/T_{1}=\frac{k_{B}T}{2\mu _{B}^{2}\hbar ^{2}\omega }\sum\limits_{i}F(Q_{i})%
\int \frac{d^{2}q}{(2\pi )^{2}}\chi ^{\prime \prime }(q,\omega \rightarrow 0)
\end{equation}%
where $Q_{i}$ stands for one peak, \ hyperfine ''tensor'' $\ F(Q)=\{A_{\perp
}+2B[\cos (Q_{x})+\cos (Q_{y})]\}^{2}$ and for $\chi ^{\prime \prime
}(q,\omega \rightarrow 0)$ we take near single  peak, say $[\pi (1-\delta
),\pi ]$
\begin{equation}
\chi ^{\prime \prime }(q,\omega )=\frac{\chi _{peak}^{\prime \prime
}(T)\omega }{\left[ 1+(x\xi _{x})^{2}+(y\xi _{y})^{2}\right] ^{2}}
\end{equation}%
where $(x,y)=(q_{x}-\pi (1-\delta );q_{y}-\pi )$ and $\xi _{x}$ and $\xi _{y}
$ are the correlation lengths in the two proper directions. After
integration the contribution from stripes with $q$ along the $x$-direction
is
\begin{equation}
1/^{63}T_{1}=\frac{k_{B}T}{\pi \mu _{B}^{2}\hbar \xi _{x}\xi _{y}}\{A_{\perp
}-2B[\cos (\pi \delta )+1]\}^{2}\chi _{peak}^{\prime \prime }
\end{equation}%
Experimentally \cite{Aeppli} $\chi _{peak}^{\prime \prime }(T)\propto T^{-2}$
and for $x=0.14$ $\ \ \ \delta =0.245\sim $1/4.\ Assuming the $T^{-1}$
dependence \cite{Aeppli} only for the one of $\xi $'s, $\xi _{x}$ and using\
for $A_{\perp }$ and $B$ the known values \cite{Zha} one obtains: $%
1/^{63}T_{1}=(4/\xi _{y})$ msec$^{-1}$. With the AF correlation length $\xi
_{y}\sim $4 this is the correct order of magnitude.

The descending dependence of the offset (Fig.3) agrees
qualitatively with the behavior of $\delta (x)$ \cite{Yamada} in
eq.(4). For a quantitative description one need to know the
$x$-dependence for $\chi _{peak}^{\prime
\prime }(T)$. Such data in the absolute units are absent yet except \cite%
{Aeppli}. In addition, with the \textit{x}-increase buckling in
the CuO$_{2}$-planes is known to decrease diminishing pinning
effects and making the local symmetry of the CuO$_{2}$-unit same
as in other materials from the class with small offset in Fig.3.
The system grows metallic with a high holes` content
\cite{31,33,34}.

Thus the only  "pseudogap" feature in the NMR data that survives
in Fig.2c is the one for YBCO 124: a change of the regime between
130 and 180 K. Similar drop in the width for the Ho-crystal field
excitations was found in \cite{Temperano} for another
124-compound, Ho-124, in about the same temperature interval. It
would be tempting to view the regime change again in the PS-terms.
However, note that the AF fluctuations cancel at the Ho-positions,
and the relaxation seen in \cite{Temperano} looks as "metallic"
Korringa law. In \cite{Temperano} DOS's, as characterizing linear
slops in T-dependencies outside the crossover interval, are almost
the same, so that whatever means the transitional area in Fig.2c
for Ho-124, it is not a simple gap opening \cite{27}. Recall that
properties of both YBCO 124 and the optimally doped LSCO (for
review see \cite{14}) are unusual in a very broad temperature
interval and are best understood in terms of the "marginal" Fermi
liquid model \cite{Varma}. Therefore we leave the origin of the
"universal " term in eq.(1) open to further discussions.

Two words regarding our choice $(\xi _{x},\xi _{y})$. We assumed anisotropic
spin correlation widths with the one along \textit{commensurate} AF
direction temperature independent.  The asymmetry is expected, although the
experimental verification remains to be done. Rapid growth of the width of
the neutron peak along modulation direction, $q_{x}$, (linear in temperature %
\cite{Aeppli}) reflects the low energy cost for magnetic excitations in LSCO
($x=0.14$) (the width from the stripes ``bending'' should have given the $%
T^{1/2}$ dependence for both $\xi $'s).

To summarize, we have found that in a temperature interval above $T_{c}$ and
below some $T^{\ast }\sim $300 K the nuclear spin relaxation for a broad
class of cuprates comes from two independent mechanisms: relaxation on
the``stripe``-like excitations that leads to a temperature independent
contribution to $1/^{63}T_{1}$ (originating due to the presence of external
doping or disorder), and an ``universal'' temperature dependent term. For La$%
_{1.86}$Sr$_{0.14}$CuO$_{4}$ we obtained a correct quantitative estimate for
the value of the first term. We argue that the whole pattern fits well the
notion of the dynamical PS into coexisting metallic and IC magnetic phases.
Experimentally, it seems that with the temperature decrease dynamical PS
acquires the static character with the IC symmetry breaking for AF phase
dictated by the competition between the lattice and the Coulomb forces.

One of the authors (L.P.G) expresses his gratitude to T.Egami, C.P.Slichter and J.Haase for
interesting discussions. The work of L.P.G. was supported by the
NHMFL through NSF cooperative agreement DMR-9527035 and the State
of Florida, that of G.B.T. through the RFBR Grant N 04-02-17137.


\begin{references}

\bibitem{1} L.P. Gor'kov and A.V. Sokol, JETP Lett. \textbf{46}, 420 (1987).

\bibitem{2} J.E. Hirsch, E. Loch \textit{et\ al.}, Phys. Rev. B \textbf{39}, 243 (1989).

\bibitem{3} J. Zaanen  \textit{et\ al.}, Phys. Rev. B \textbf{40}, 7391 (1989).

\bibitem{4} V.J. Emery \textit{et\ al.}, Phys. Rev. Lett. \textbf{64}, 475 (1990).

\bibitem{5} M. Grilli \textit{et\ al.}, Phys. Rev. Lett. \textbf{67}, 259 (1991).

\bibitem{6} T. Egami and S. J. L. Billinge, in Physical Properties of High-Temperature
Superconductors V, edited by D. M. Ginsberg (World Scientific, Singapore,
1996), 265.


\bibitem{8} S.H. Pan \textit{et\ al.}, Nature \textbf{413}, 282 (2001).

\bibitem{Howald} C. Howald \textit{et al., }Phys.Rev. B \textbf{67}, 014533 (2003).

\bibitem{Salcola} M.I.Salkola \textit{et\ al.}, Phys. Rev. Lett. \textbf{77}, 155 (1996);

R.S.Markiewicz \textit{et\ al.}, Phys. Rev. B \textbf{65},
064520 (2002).

\bibitem{9} J.M. Tranquada \textit{et\ al.}, Nature \textbf{375}, 561 (1995).




\bibitem{14} T. Timusk and B. Statt,  Rep. Prog. Phys. \textbf{62}, 61 (1999).

\bibitem{15} J.L. Tallon and J.M. Loram, Physica C \textbf{349}, 53 (2001).


\bibitem{17} L.P. Gor'kov, Journ. Supercond. \textbf{14}, 365 (2001).

\bibitem{Chakra} S. Chakravarty \textit{et\ al.}, Phys. Rev. B \textbf{63}, 094503 (2001).

\bibitem{Kam} A. Kaminsky \textit{et\ al.}, Nature \textbf{416}, 610 (2002).

\bibitem{18} G. Teitel'baum \textit{et\ al.}, Phys. Rev. Lett. \textbf{84}, 2949
(2000).

\bibitem{19} G. Teitel'baum \textit{et\ al.}, Phys. Rev. B \textbf{63}, 020507(R)
(2001).






\bibitem{31} S. Uchida \textit{et\ al.}, Physica C \textbf{162-164,} 1677 (1989).


\bibitem{33} T. Nishikawa \textit{et\ al.}, J. Phys. Soc. Jpn. \textbf{62}, 2568 (1993).

\bibitem{34} F. Balakirev \textit{et\ al.}, Nature \textbf{424}, 912 (2003).

\bibitem{27} G.V.M. Williams \textit{et\ al.}, Phys. Rev. B \textbf{58}, 15053 (1998).


\bibitem{Schmallian} J. Schmalian \textit{et al}., Phys.Rev. B \textbf{60}, 667 (1999).

\bibitem{39} S.Oshugi  \textit{et\ al.}, J. Phys. Soc. Jpn. \textbf{63}, 700 (1994).

\bibitem{Imai} T. Imai \textit{et al.}, Phys. Rev. Lett. \textbf{70}, 1002 (1993).

\bibitem{Singer} P. M. Singer \textit{et\ al.}, Phys. Rev. Lett. \textbf{88}, 187601
(2002).

\bibitem{26} M. Takigawa \textit{et\ al}., Phys. Rev. B \textbf{43}, 247 (1991).

\bibitem{Walstedt} R.E. Walstedt \textit{et al}., Phys.Rev. B \textbf{\ 44}, 7760 (1991).





\bibitem{G} S.W. Cheong \textit{et al}.,  Phys. Rev. Lett. \textbf{67}, 1791 (1991); H.A.
Mook \textit{et\ al.},  Nature \textbf{395}, 580 (1998); M. Arai
\textit{et al.}, Phys. Rev. Lett. \textbf{83}, 608 (1999);  A.
Bianconi, Int. J.Mod.Phys. B \textbf{14}, 3289 (2000); P. Dai
\textit{et\ al.}, Phys. Rev. B \textbf{63}, 054525 (2001).

\bibitem{Barzykin} V. Barzykin \textit{et al}., Phys. Rev. B \textbf{50}, 16052 (1994).

\bibitem{Crawford} M.K. Crawford \textit{et al}., Phys. Rev. B \textbf{44}, 7749 (1991).

\bibitem{Tranquada1} J.M. Tranquada \textit{et al}., Phys. Rev.Lett. \textbf{78}, 338 (1997).

\bibitem{Tranquada2} J.M. Tranquada \textit{et al}., Phys. Rev. B \textbf{54}, 7489 (1996).

\bibitem{Tranquada3} J.M. Tranquada \textit{et al}., Phys. Rev. B \textbf{59}, 14712 (1999).

\bibitem{Fujita} M. Fujita \textit{et al}., Phys.Rev. B \textbf{65}, 064505 (1991).

\bibitem{Lake} B. Lake \textit{et al}., Nature \textbf{415}, 299 (2002).

\bibitem{Niedermeier} Ch. Niedermayer \textit{et al}., Phys. Rev.Lett. \textbf{80}, 3843 (1998).

\bibitem{Yamada} K. Yamada \textit{et\ al.}, Phys. Rev. B \textbf{57}, 6165 (1998).

\bibitem{Aeppli} H. Aeppli \textit{et al}., Science \textbf{278}, 1432 (1997).

\bibitem{Zha} Y. Zha \textit{et al}., Phys. Rev. B \textbf{54}, 7561(1996).

\bibitem{Temperano} D. Rubio Temprano \textit{et\ al.}, Phys. Rev. Lett. \textbf{84},
1990 (2000).

\bibitem{Varma} C. M. Varma \textit{et\ al.}, Phys. Rev. Lett. \textbf{63}, 1996
(1989).











\end{references}
\end{document}